# A Simple and Efficient Equivariant Message Passing Neural Network Model for Non-Local Potential Energy Surface

*Yibin Wu[1], Junfan Xia[1], Yaolong Zhang[3]\*, Bin Jiang[2]\*.*

1. Heifei National Laboratory for Physical Science at the Microscale, Department of Chemical Physics, University of Science and Technology of China, Hefei, Anhui, 230026, China.

2. Key Laboratory of Precision and Intelligent Chemistry, Department of Chemical Physics, University of Science and Technology of China, Hefei, Anhui 230026, China; ORCID: https://orcid.org/0000-0003-2696-5436; Email: bjiangch@ustc.edu.cn

3. Department of Chemistry and Chemical Biology, Center for Computational Chemistry, University of New Mexico, Albuquerque, New Mexico 87131, USA; Email: ylzhangch@unm.edu

**ABSTRACT.**

Machine learning potentials have become increasingly successful in atomistic simulations. Many of these potentials are based on an atomistic representation in a local environment, but an efficient description of non-local interactions that exceed a common local environment remains a challenge. Herein, we propose a simple and efficient equivariant model, EquiREANN, to effectively represent non-local potential energy surface. It relies on a physically inspired message passing framework, where the fundamental descriptors are linear combination of atomic orbitals, while both invariant orbital coefficients and the equivariant orbital functions are iteratively updated. We demonstrate that this EquiREANN model is able to describe the subtle potential energy variation due to the non-local structural change with high accuracy and little extra computational cost than an invariant message passing model. Our work offers a generalized approach to create equivariant message passing adaptations of other advanced local many-body descriptors.

Atomistic simulations are powerful computational tools for understanding molecules, chemical reactions, and materials. The reliability of molecular simulations largely relies on the accuracy of the potential energy surface (PES) used. In recent years, machine learning (ML) methodologies have been extensively developed to accurately represent PESs from *ab initio* data points[1-24], which have been applied into a diversity of areas in chemistry[25-38]. These machine-learned potentials (MLPs) have demonstrated their superior accuracy than empirical force fields and much lower costs than direct *ab initio* molecular dynamics simulations.

A very successful family of MLPs are based on an atom-based local representation of the PES, which was originally introduced by Behler and Parrinello in their high-dimensional neural network approach.[2] This type of PES decomposes the total energy of the system into atomic energies which are dependent on each atomic environment within a cutoff radius. The local atomic environment is then described by a family of many-body atomic descriptors thus preserve the translational, rotational, and permutational invariance of the potential energy. Truncating atomic interactions within a local environment makes the computational cost and scaling more favorable in high-dimensional systems than a global representation. However, non-local interactions beyond the cutoff radius are not well captured.[28]

More recently, message passing neural network (MPNN) based approaches, have become increasingly popular to learn the atom-based representation of the PES in an end-to-end way. [8, 11, 15, 17-19, 21, 22] These approaches also take advantage of the atomic energy decomposition and represent the atomic environment by iteratively passing messages between nodes (*e.g.*, atoms) connected by edges (*e.g.*, interatomic distances and/or angles) in the context of molecular graphs. After several iterations, this scheme can create a representation that indirectly carries information about atoms from outside of the cutoff radius and encodes high-order many-body interactions.[17]

As a result, MPNN methods can generally describe non-local interactions or in other words extend the effective cutoff radius (equal to the cutoff multiplied by the number of message passing) of the atomic environment.

However, the effectiveness of describing the atomic environment and non-local features in an MPNN model may depend on the complexity of the message being passed. The first generation of MPNN architectures, like SchNet[8] and PhysNet[11], were built on message-passing of two-body functions. It was later realized that including three-body (or even higher-order body) interactions in the message-passing can better distinguish certain atomic structures, which is crucial to increase the representability of the MPNN model, *e.g.*, SpookyNet[15]. Interestingly, three-body based MPNN models have been also effectively achieved in a physically inspired way by making some atom-wise coefficients of atomic descriptors dependent on their respective neighboring atomic environment, such as the recursively embedded atom neural network (REANN)[17] and AIMNet[19] models. These models are often referred to as invariant MPNN models, in which message-passing functions that are invariant with respect to symmetry operations. More recently, various equivariant MPNN architectures were proposed by passing equivariant features to better capture interactions depending on the relative orientation of neighboring atoms.[18, 20-22]

A special category of systems that highlights the non-locality of the PES are cumulenes, hydrocarbons of the form $C_nH_4$ with $n$ - 1 cumulative double bonds.[28] Because of the uninterrupted conjugated bonding between neighboring carbon atoms, these molecules possess a rigid linear backbone with the two terminal methylene groups. Consequently, the potential energy varies with the dihedral angle specifying the relative orientation of the $CH_2$ groups and its variation depends the length of the carbon chain, even when $n$ is large. These systems provide stringent tests against atom-based MLPs for a proper description of the global structure. Indeed, it was found that even with a fairly large cutoff radius (*e.g.* 12.0 Å), the SchNet and PhysNet models failed to reproduce any energy dependence on the dihedral angle when the carbon chain becomes long (*e.g.*, $n = 9$)[28]. Moreover, neither an inherently global kernel-based symmetrized gradient domain machine learning (sGDML[10]) model nor a SpookyNet[15] model that explicitly includes non-local corrections can correctly describe the energy profile at a finite cutoff[22]. Very recently, a self-attention based model, So3krates, was specially designed to efficiently incorporate non-local interactions in the spherical harmonic coordinate (SPHC) space, where seemingly distant atoms in the Cartesian coordinate space can be correlated in the SPHC space.[22] However, this particular treatment aims to introduce correlations between similar atomic structures in SPHC space rather than introducing general long-range interactions. In addition, additional operations in SPHC space would reduce the efficiency of the model[22].

In this work, we develop a simple and efficient equivariant implementation of REANN model. In this equivariant model, the fundamental atomic feature remains given by the square of the linear combination of atomic orbitals of neighboring atoms. However, not only invariant orbital coefficients but also the equivariant orbitals are iteratively updated, which implicitly contains the spatial information of neighboring atoms. We demonstrate that this equivariant model more efficiently propagates the directional information through the linear carbon chain than the invariant counterpart, thus is able to describe the PESs of cumulenes with high accuracy and little extra computational cost.

Let us begin with the original EANN model. In the EANN framework, the local environment is characterized by embedded atom density (EAD) descriptors[12]. These physically inspired EAD descriptors are constructed from the square of the linear combination of atomic orbitals situated at neighboring atoms. Consequently, an EAD component centered at atom $i$, $\rho_i$, can be expressed as follows,

$$\rho_i = \sum_{l=0}^{L} \sum_{l_x,l_y,l_z}^{l_x+l_y+l_z=l} \frac{l!}{l_x!\,l_y!\,l_z!} \left( \sum_{m}^{N_\varphi} d_m \sum_{j\neq i}^{N_c} c_j \varphi_m^{l_x l_y l_z}(\hat{\mathbf{r}}_{ij}) f_c(r_{ij}) \right)^2, \quad (1)$$

Here, $\varphi_m^{l_x l_y l_z}(\hat{\mathbf{r}}_{ij})$ is a primitive Gaussian-type orbital (GTO),

$$\varphi_m^{l_x l_y l_z}(\hat{\mathbf{r}}_{ij}) = (x_{ij})^{l_x}(y_{ij})^{l_y}(z_{ij})^{l_z} \exp[-\alpha_m(r_{ij}-r_m)^2], \quad (2)$$

where $\hat{\mathbf{r}}_{ij} = (x_i - x_j, y_i - y_j, z_i - z_j)$ represents the relative position vector of the neighboring atom $j$ pointing to the central atom $i$, the angular momentum $l$ and its projections ($l_x$, $l_y$, and $l_z$) specify the orientation of the GTO, $\alpha_m$ and $r_m$ are learnable parameters determining the shape of the GTO, $d_m$ and $c_j$ are learnable contraction and orbital coefficients, respectively. For numerical efficiency, the summation order is exchanged, as the number of neighboring atoms ($N_c$) is often larger than the number of primitive GTOs. A radial cutoff function $f_c(r_{ij})$ ensures the interactions smoothly decay to zero at a cutoff radius ($r_c$). The number of contracted orbitals ($N_\varphi$) and maximum angular momentum ($L$) determine the size of the EAD feature vector and its representability to the local environment.

A purely local EAD descriptor only includes at most three-body interactions[12]. To introduce higher-order interactions, $c_j$ can be made dependent on the local environment of atom $j$, expressed as the output of an NN based on EAD features centered at atom $j$ in a message passing form[17, 39],

$$c_j^t = g^{t-1}(\rho_j^{t-1}(\mathbf{c}_j^{t-1}, \hat{\mathbf{r}})). \tag{3}$$

where $g^{t-1}$ denotes an NN at ($t-1$)th message-passing layer with the EAD feature $\rho_j^{t-1}$ being its input. After $T$ rounds of message-passing, the last NN outputs the atomic energy, $E_i$, whose sum gives the total potential energy $E$. In this REANN framework, only the scalar orbital coefficients are iteratively updated, which thus corresponds to an invariant MPNN architecture.[36]

To illustrate the difficulty of representing the non-local effect by invariant MPNNs like REANN, we take the simplest ethene molecule as an example in Fig. 1. Here, a small cutoff radius of 1.5 Å is used to highlight the problem and the corresponding local environments of the C1 and C2 atoms are presented by light blue and light red circles. The two conformers in Fig. 1(a) differ only in the H-C-C-H dihedral angle and the enclosed angles of H1(H2)-C1(C2)-H3(H4), which are certainly not included in any local environment. While the C1 center can acquire some extra correlation beyond its local environment through message-passing of the orbital coefficients at the C2 atom, such like the three-body interaction C1-C2-H3(H4), it cannot establish a direct correlation between H1(H2) and H3(H4), regardless of the number of message-passing iterations. This limitation renders the two conformers indistinguishable in the REANN model.

One brute-force solution to this issue is to enlarge the cutoff and include all H atoms within the local environment of at least one carbon atom. However, this often leads to a significant rise in computational time. Alternatively, we can pass not only the *scalar* orbital coefficient but also the *tensorial* linearly combined orbital function, namely

$$\psi^0_{i,l_x,l_y,l_z,m} = \sum_{j\neq i}^{N_c} \left( c_j^0 \varphi_m^{l_x l_y l_z}(\hat{\mathbf{r}}_{ij}) \right), \tag{4}$$

$$\psi^t_{i,l_x,l_y,l_z,m} = \sum_{j\neq i}^{N_c} \left( c_j^t \varphi_m^{l_x l_y l_z}(\hat{\mathbf{r}}_{ij}) + \psi^{t-1}_{i,l_x,l_y,l_z,m} f_c(r_{ij}) \right), \tag{5}$$

where for $L>0$, $\psi^{t-1}_{i,l_x,l_y,l_z,m}$ is an equivariant feature as the GTO equivariantly transforms with respect to the spatial position of the neighboring atom relative to the central atom. The invariant EAD feature can be obtained using the same form as Eq. (1),

$$\rho_i^t = \sum_{l=0}^{L} \sum_{l_x,l_y,l_z}^{l_x+l_y+l_z=l} \frac{l!}{l_x! l_y! l_z!} \left( \sum_m^{N_\varphi} d_m \psi^t_{i,l_x,l_y,l_z,m} \right)^2. \tag{6}$$

This equivariant REANN model is referred as EquiREANN hereafter, whose architecture is illustrated in Fig. 2. Its superiority over the invariant REANN model is interpreted in Fig. 1(b) for the same ethene molecule. For better visualization, we take the $L=1$ case as an example, where the linearly combined orbital function at the C2 center, $\psi^0_{C2}$, is a vectorial function, which incorporates the directional information of all neighbor atoms relative to C2 and depends on the combinative orientation of $\hat{\mathbf{r}}_{C2C1}$, $\hat{\mathbf{r}}_{C2H3}$, and $\hat{\mathbf{r}}_{C2H4}$. This message function is rotationally equivariant in the local environment of C2. Through message-passing from C2 to C1, $\psi^0_{C2}$ will merge into the EAD feature of the C1 center, by which the two structures become distinguishable. More explicitly, we can expand an EAD component at the C1 center after message-passing once from C2 (where the pre-factor, the Gaussian functions, and contraction coefficients are ignored for simplicity),

$$\begin{aligned}\rho_{C1} &= \left\| c^1_{C1H1}\hat{\mathbf{r}}_{C1H1} + c^1_{C1H2}\hat{\mathbf{r}}_{C1H2} + (c^1_{C1C2} + c^0_{C1C2})\hat{\mathbf{r}}_{C1C2} + c^0_{C2H3}\hat{\mathbf{r}}_{C2H3} + c^0_{C2H4}\hat{\mathbf{r}}_{C2H4} \right\|^2 \\ &= \left\| c^1_{C1H1}\hat{\mathbf{r}}_{C1H1} + c^1_{C1H2}\hat{\mathbf{r}}_{C1H2} + (c^1_{C1C2} + c^0_{C1C2} - 1)\hat{\mathbf{r}}_{C1C2} + c^0_{C2H3}\hat{\mathbf{r}}_{C2H3} + (c^0_{C2H4} - 1)\hat{\mathbf{r}}_{C2H4} + \hat{\mathbf{r}}_{C1H4} \right\|^2 . \quad (7) \\ &= c^1_{C1H1}\langle \hat{\mathbf{r}}_{C1H1} \cdot \hat{\mathbf{r}}_{C1H4}\rangle + c^1_{C1H2}\langle \hat{\mathbf{r}}_{C1H2} \cdot \hat{\mathbf{r}}_{C1H4}\rangle + \cdots \end{aligned}$$

Note that these vector products associated with C1, *e.g.* the first two terms, in the final expression, generate the angular correlations of H1-C1-H4 and H2-C1-H4, which are essential to identify the structural difference caused by rotating the terminal $CH_2$ group.

Numerical tests have been performed to validate the EquiREANN model in representing the non-locality of the PESs of cumulenes. To this end, we constructed a dataset consisting of 9100 configurations for each cumulene molecule ($C_nH_4$, $n$ = 2-9), with their bond lengths and bond angles randomly deviated from respective equilibrium geometries. These structures cover the following ranges of C-H bond length (1.05 Å to 1.15 Å), C-C bond length (1.25 Å to 1.35 Å), C-C-C angle (170° to 190°), and H-C-C angle (110° to 130°). As our purpose is to verify the model's capability to capture the non-local energy dependence on the dihedral angle, the energies and forces of these configurations were calculated at the level of density functional theory (DFT) using the B3LYP functional and a 6-31G basis set.[40, 41] Unless otherwise specified, the PES training was done for each individual molecule, in which the dataset was divided into a training set (90 % of the data points) and a validation set (10 % of the data points). An initial learning rate of 0.005 with a decay factor of 0.5 was employed to optimize the parameters of the model and training was considered converged when the learning rate decreased below $5\times10^{-5}$. A batch size of 64 was used for training. A two hidden layer 16×16 NN architecture was used in each iteration, $N_\varphi$ and $L$ was set to 8 and 2. All other parameters are learned in an end-to-end fashion.

Fig. 3 shows the energy profiles predicted by the REANN[17] and EquiREANN methods, as well as the reference data calculated by DFT, as a function of the dihedral angle for three representative cumulenes. Also shown are the results of some other MLP models[8, 10] for the same system along with their reference data[28] calculated by a semi-empirical MNDO method (*i.e.*, modified neglect of diatomic overlap)[42]. We adopt the same hyper-parameter setting for REANN and EquiREANN models, with $r_c$ = 6.0 Å and $T$ = 2, respectively. For $C_5H_4$ and $C_7H_4$, REANN can capture the energy variation because all H atoms are included in the local environment of the middle carbon atom, marked with an orange ×. In this case, the hydrogen atoms at opposite ends are directly correlated by the local three-body EAD descriptors, while higher-order interactions are introduced by passing the scalar orbital coefficients. Note, for $C_7H_4$, that REANN does not perfectly match the DFT reference energy, which may be improved by increasing the number of message-passing layers, as demonstrated in Ref. 17. For $C_9H_4$, the terminal H atoms are outside the cutoff sphere of any atom so that REANN cannot correlate them via invariant message-passing and predicts a completely constant energy as the dihedral angle varies. In comparison, the EquiREANN model describes the energy variation very well for all cumulenes. The key for its success is the equivariant message-passing that effectively correlates the H atoms at opposite ends.

It is worthwhile to note that for two-body feature based MPNN models such as SchNet[8], only these atoms with interatomic distances smaller than $r_c$ can be explicitly correlated[17]. Consequently, even all atoms are included in the cutoff sphere of the middle carbon atom with $r_c$ = 6.0 Å, for $C_5H_4$ and $C_7H_4$, SchNet cannot describe the energy variation accurately. Increasing $r_c$ to 12.0 Å solves the problem for $C_5H_4$ and $C_7H_4$, as $r_c$ is long enough to enable an H-centered local environment to include the H atoms at the other end. However, for $C_9H_4$, these terminal H atoms are too distant from each other and remain uncorrelated, leading to a constant energy prediction. Interestingly, the global sGDML[10] model incrementally underestimates the extent of energy variation as the length of carbon chain increases. This is possibly because the variation of the distance between these terminal H atoms with their relative orientation is too weak, especially for a long carbon chain, which can be hardly captured in the global descriptor of sGDML.

To further clarify the effectiveness of the EquiREANN model, we test the model performance using a small cutoff of $r_c$ = 3.0 Å and an increasing number of message-passing layers up to $T$ = 2 for $C_5H_4$. The corresponding results of REANN, EquiREANN, as well as a popular equivariant MPNN model, namely NequIP[20], are compared in Fig. 4. For $T$ = 0, the cutoff is too small and none of these models would work. Increasing $T$ amounts to increasing the so-called effective cutoff sphere or receptive field of each atom center that allows more atoms to be indirectly correlated. As analyzed in Fig. 1, however, the invariant message-passing scheme in the REANN model cannot correlate the opposite terminal H atoms that are outside the small cutoff sphere. This results in a constant energy predicted by REANN for any $T$. In comparison, EquiREANN ($T$ = 1) can indirectly correlate all H atoms in the middle carbon atom-centered features, predicting different atomic energies for different structures and allowing the model to capture the change of the total energy reasonably well. Increasing $T$ to 2 includes more high-order interactions to the model and the terminal H-H correlations in other carbon atom-centered features, thus leading to a perfect agreement with the reference data. On the other hand, NequIP characterizes the nonlocal correlation by the sequential tensor product of nearest neighbor equivariant features in the equivariant interaction block. While this scheme accumulates more high-order correlations in an atomic feature, yielding higher distinguishability for complicated atomic structures than many others[20], it cannot correlate atoms in the opposite directions of the message-passing chain. In this case, the H atoms at opposite ends of the cumulene can only be correlated if the effective cutoff radius exceeds the length of the carbon chain, namely by NequIP ($T = 2$, or three interaction blocks), in a way that the message-passing goes from the left-end carbon atom to the right-end one. In this regard, NequIP is less efficient than EquiREANN for incorporating non-local effects. This analysis also explains the finding in Ref. [22], where the NequIP model with $r_c = 2.5$ Å requires six iterations of message-passing to barely describe the energy variation of $C_9H_4$.

An important feature of EquiREANN is that its equivariant adaption introduces very simple operations and adds little computational cost compared to REANN. In practice, the comparison of training times for $C_7H_4$ with the same hyperparameters used for REANN (6.2 s/Epoch) and EquiREANN (5.9 s/Epoch) reveals that the additional computational cost is negligible. Moreover, the training time for NequIP is 16.7 seconds/Epoch with similar hyperparameter settings and the same batch size. In this respect, EquiREANN is an efficient implementation of equivariant MPNNs. Notably, So3krates captures non-local features in cumulenes very effectively by introducing SPHC to correlate hydrogen atoms at opposite ends in the SPHC space. However, this special scheme significantly increases computational cost, as discussed by the authors themselves of So3krates[22]. However, we find that the publicly available So3krates code contains no implementation for the SPHC treatment, which prevent us to perform a direct comparison at this point.

Summarizing, we demonstrate in this work that how to adapt an invariant MPNN to an equivariant MPNN in a simple and efficient way. Our method keeps the simplicity of the physically inspired REANN framework and update the tensorial orbital functions along with the scalar orbital coefficients in the message-passing process, leading to an EquiREANN model. Numerical tests on cumulene molecules validate that EquiREANN largely improves the description of non-local PESs compared to invariant REANN models. Additionally, this message construction scheme has a minimal impact on the computational cost, as the number of parameters remains unchanged. This concept can be readily extended to create equivariant adaptations of other advanced local many-body descriptors, preserving their inherent structures. This strategy holds promise for developing more accurate and efficient machine learning potentials, particularly for complex systems where non-local interactions play a significant role.

**ACKNOWLEDGMENTS**

This work is supported by the Strategic Priority Research Program of the Chinese Academy of Science (XDB0450101), CAS Project for Young Scientists in Basic Research (YSBR-005), National Natural Science Foundation of China (22325304, 22221003 and 22033007). We acknowledge the Supercomputing Center of USTC, Hefei Advanced Computing Center, Beijing PARATERA Tech CO., Ltd for providing high-performance computing service.

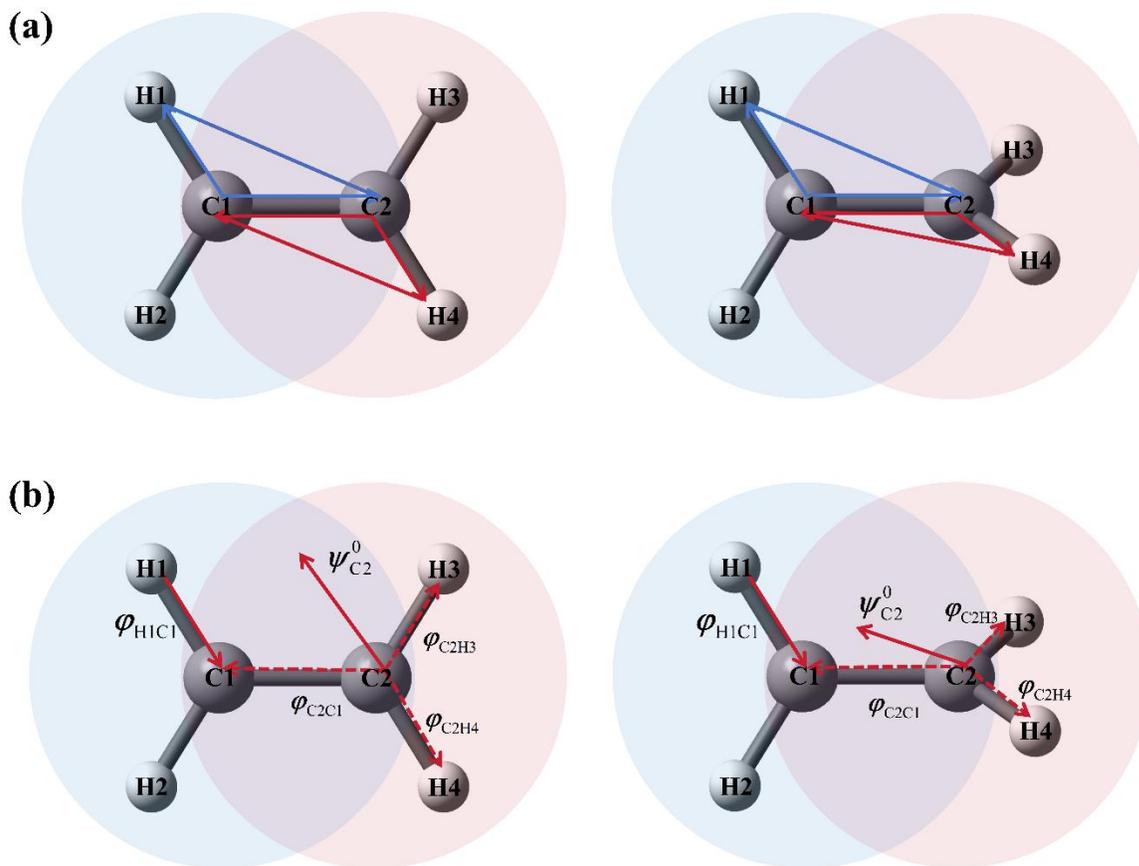

**Figure 1.** Schematic diagram of the message-passing process for (a) REANN and (b) EquiREANN, applied to two different ethene structures. Light blue and red circles denote the local environments of C1 and C2 centers, for which message is passed from C2 to C1. In panel (a), the blue and red solid lines represent the H1-C1-C2 correlation of C1-center and the C1-C2-H4 correlation of C2-center, respectively. In panel (b), the additional equivariant message, $\psi_{C_2}^0$, is generated by a linear combination of GTOs (dashed lines) of C2-center, which reorients as the H1-C1-C2-H4 dihedral angle rotates.

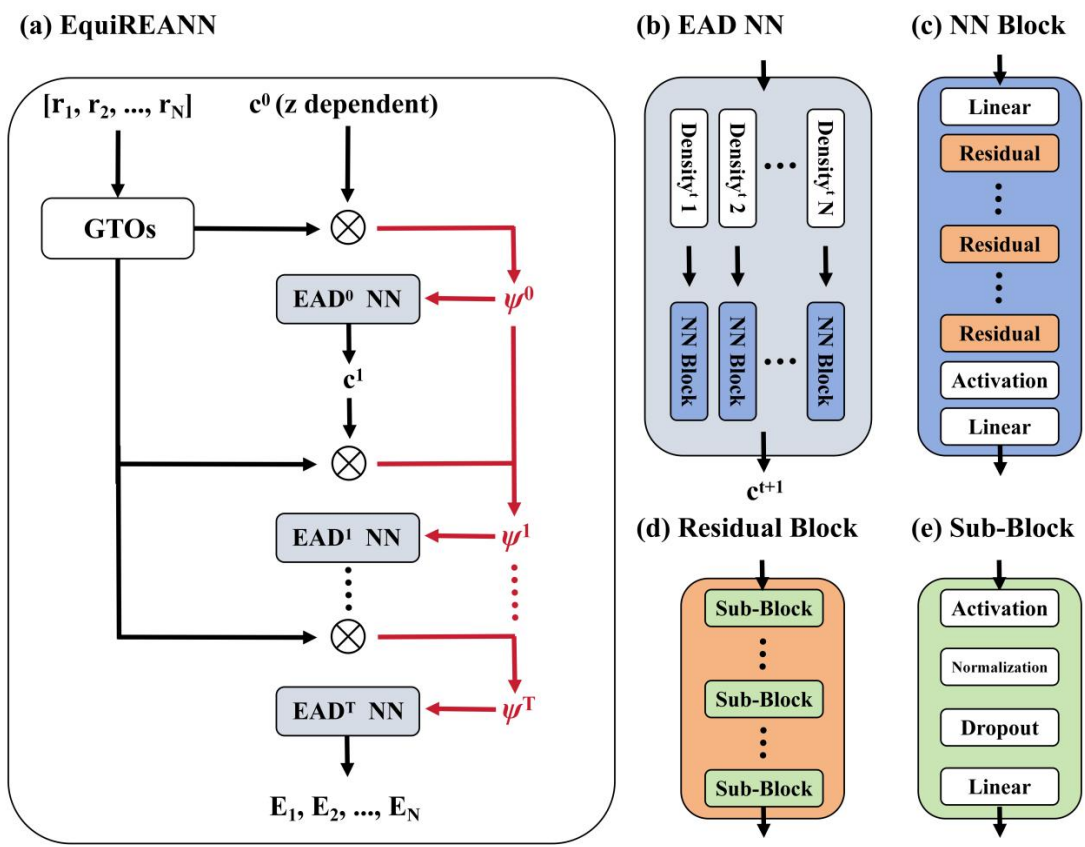

**Figure 2.** Schematic workflow of the EquiREANN architecture. The structure of the NN module is the same as that of the REANN package[39].

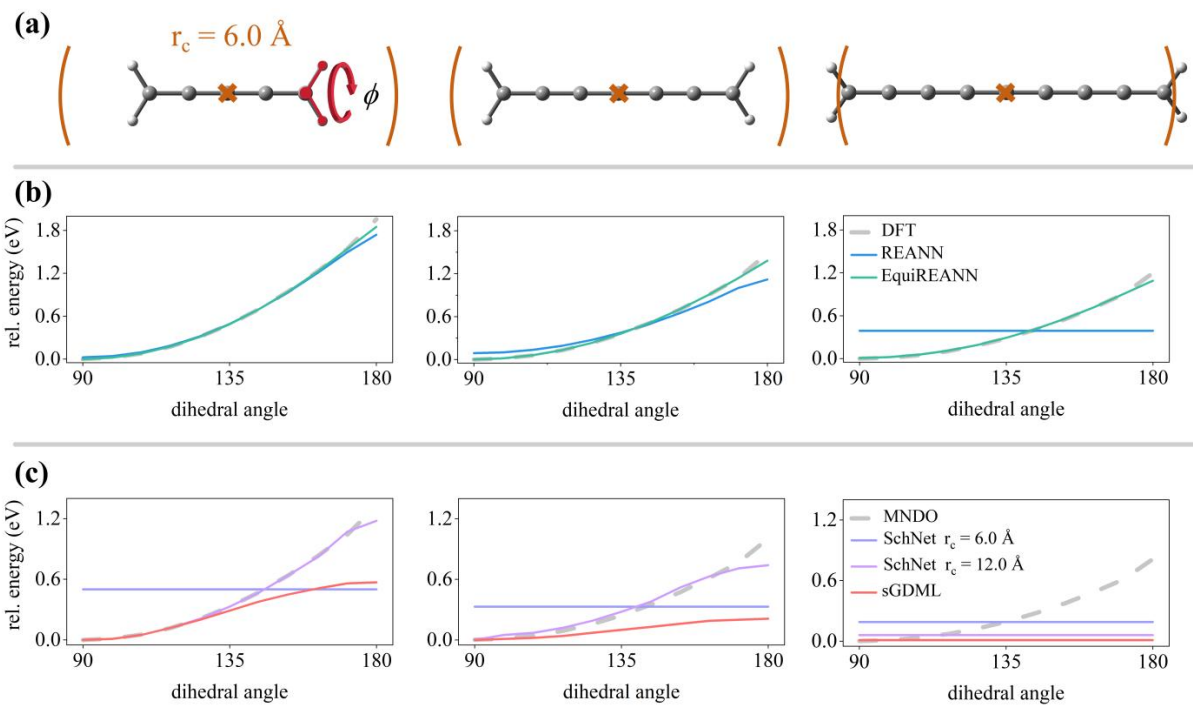

**Figure 3.** (a) Schematic diagrams of representative cumulenes, $C_5H_4$, $C_7H_4$, and $C_9H_4$, along with their respective cutoff spheres centered at the middle carbon atom. (b-c) Energy profiles as a function of the dihedral angle ($\phi$) in these cumulenes calculated with REANN, EquiREANN, SchNet[8] and sGDML[10]. The cutoff is set to 6.0 Å if not mentioned otherwise. Note that the reference data have been generated with different methods (DFT or MNDO) for (b) and (c), respectively, which however does not affect the comparison of the energy trend.

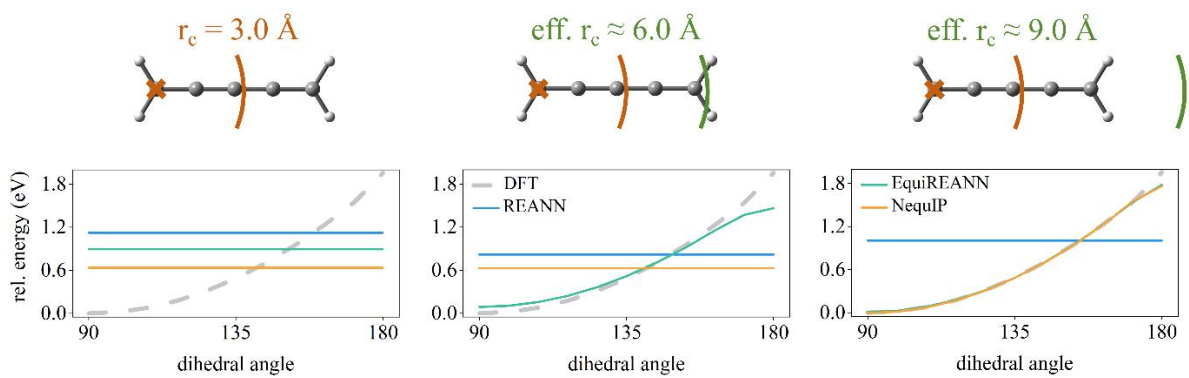

**Figure 4.** Energy profiles as a function of the H-C-C-H dihedral angle in $C_5H_4$ with message passing times increasing from $T$=0 to 2 (from left to right panels) obtained by REANN[17], EquiREANN, and NequIP[20] models, in comparison with DFT reference data. In all models, the cutoff radius is 3.0 Å and illustrated centered in terminal carbon atom for a better illustration. The effective cutoff is proportionally extended as $T$ increases, which is estimated and displayed taking the left-terminal carbon atom as the center for reference.

**References**

(1) Manzhos, S.; Carrington Jr., T. Using neural networks to represent potential surfaces as sums of products. *J. Chem. Phys.* **2006**, *125*, 194105.

(2) Behler, J.; Parrinello, M. Generalized neural-network representation of high-dimensional potential-energy surfaces. *Phys. Rev. Lett.* **2007**, *98*, 146401.

(3) Bartók, A. P.; Payne, M. C.; Kondor, R.; Csányi, G. Gaussian approximation potentials: The accuracy of quantum mechanics, without the electrons. *Phys. Rev. Lett.* **2010**, *104* (13), 136403.

(4) Jiang, B.; Guo, H. Permutation invariant polynomial neural network approach to fitting potential energy surfaces. *J. Chem. Phys.* **2013**, *139* (5), 054112.

(5) Shao, K.; Chen, J.; Zhao, Z.; Zhang, D. H. Communication: Fitting potential energy surfaces with fundamental invariant neural network. *J. Chem. Phys.* **2016**, *145* (7), 071101.

(6) Shapeev, A. V. Moment Tensor Potentials: A Class of Systematically Improvable Interatomic Potentials. *Multiscale Model. Sim.* **2016**, *14* (3), 1153-1173.

(7) Yao, K.; Herr, J. E.; Toth, David W.; McKintyre, R.; Parkhill, J. The TensorMol-0.1 model chemistry: a neural network augmented with long-range physics. *Chem. Sci.* **2018**, *9* (8), 2261-2269.

(8) Schütt, K. T.; Sauceda, H. E.; Kindermans, P.-J.; Tkatchenko, A.; Müller, K.-R. SchNet – A deep learning architecture for molecules and materials. *J. Chem. Phys.* **2018**, *148* (24), 241722.

(9) Zhang, L.; Han, J.; Wang, H.; Car, R.; E, W. Deep Potential Molecular Dynamics: A Scalable Model with the Accuracy of Quantum Mechanics. *Phys. Rev. Lett.* **2018**, *120* (14), 143001.

(10) Chmiela, S.; Sauceda, H. E.; Müller, K.-R.; Tkatchenko, A. Towards exact molecular dynamics simulations with machine-learned force fields. *Nat. Commun.* **2018**, *9* (1), 3887.

(11) Unke, O. T.; Meuwly, M. PhysNet: A Neural Network for Predicting Energies, Forces, Dipole Moments, and Partial Charges. *J. Chem. Theory Comput.* **2019**, *15* (6), 3678-3693.

(12) Zhang, Y.; Hu, C.; Jiang, B. Embedded atom neural network potentials: Efficient and accurate machine learning with a physically inspired representation. *J. Phys. Chem. Lett.* **2019**, *10* (17), 4962-4967.

(13) Drautz, R. Atomic cluster expansion for accurate and transferable interatomic potentials. *Phys. Rev. B* **2019**, *99* (1), 014104.

(14) Zaverkin, V.; Kästner, J. Gaussian Moments as Physically Inspired Molecular Descriptors for Accurate and Scalable Machine Learning Potentials. *J. Chem. Theory Comput.* **2020**, *16* (8), 5410-5421.

(15) Unke, O. T.; Chmiela, S.; Gastegger, M.; Schütt, K. T.; Sauceda, H. E.; Müller, K.-R. SpookyNet: Learning force fields with electronic degrees of freedom and nonlocal effects. *Nat. Commun.* **2021**, *12* (1), 7273.

(16) Dral, P. O.; Ge, F.; Xue, B.-X.; Hou, Y.-F.; Pinheiro, M.; Huang, J.; Barbatti, M. MLatom 2: An Integrative Platform for Atomistic Machine Learning. *Top. Curr. Chem.* **2021**, *379* (4), 27.

(17) Zhang, Y.; Xia, J.; Jiang, B. Physically motivated recursively embedded atom neural networks: Incorporating local completeness and nonlocality. *Phys. Rev. Lett.* **2021**, *127* (15), 156002.

(18) Schütt, K.; Unke, O.; Gastegger, M. Equivariant message passing for the prediction of tensorial properties and molecular spectra. In *Proceedings of the 38th International Conference on Machine Learning*, Proceedings of Machine Learning Research; 2021.

(19) Zubatyuk, R.; Smith Justin, S.; Leszczynski, J.; Isayev, O. Accurate and transferable multitask prediction of chemical properties with an atoms-in-molecules neural network. *Sci. Adv.* **2021**, *5* (8), eaav6490.

(20) Batzner, S.; Musaelian, A.; Sun, L.; Geiger, M.; Mailoa, J. P.; Kornbluth, M.; Molinari, N.; Smidt, T. E.; Kozinsky, B. E(3)-equivariant graph neural networks for data-efficient and accurate interatomic potentials. *Nat. Commun.* **2022**, *13* (1), 2453.

(21) Batatia, I.; Kovacs, D. P.; Simm, G.; Ortner, C.; Csanyi, G. MACE: Higher Order Equivariant Message Passing Neural Networks for Fast and Accurate Force Fields. In *Advances in Neural Information Processing Systems*, 2022.

(22) Thorben Frank, J.; Unke, O. T.; Müller, K.-R. So3krates: Equivariant attention for interactions on arbitrary length-scales in molecular systems. In *Advances in Neural Information Processing Systems*, 2022.

(23) Zhang, Y.; Jiang, B. Universal machine learning for the response of atomistic systems to external fields. *Nat. Commun.* **2023**, *14* (1), 6424.

(24) Song, Z.; Han, J.; Henkelman, G.; Li, L. Charge-Optimized Electrostatic Interaction Atom-Centered Neural Network Algorithm. *J. Chem. Theory Comput.* **2024**, *20* (5), 2088-2097.

(25) Smith, J. S.; Isayev, O.; Roitberg, A. E. ANI-1: an extensible neural network potential with DFT accuracy at force field computational cost. *Chem. Sci.* **2017**, *8* (4), 3192-3203, 10.1039/C6SC05720A.

(26) Smith, J. S.; Nebgen, B. T.; Zubatyuk, R.; Lubbers, N.; Devereux, C.; Barros, K.; Tretiak, S.; Isayev, O.; Roitberg, A. E. Approaching coupled cluster accuracy with a general-purpose neural network potential through transfer learning. *Nat. Commun.* **2019**, *10* (1), 2903.

(27) Zeng, J.; Cao, L.; Xu, M.; Zhu, T.; Zhang, J. Z. H. Complex reaction processes in combustion unraveled by neural network-based molecular dynamics simulation. *Nat. Commun.* **2020**, *11* (1), 5713.

(28) Unke, O. T.; Chmiela, S.; Sauceda, H. E.; Gastegger, M.; Poltavsky, I.; Schütt, K. T.; Tkatchenko, A.; Müller, K.-R. Machine Learning Force Fields. *Chem. Rev.* **2021**, *121* (16), 10142-10186.

(29) Musil, F.; Grisafi, A.; Bartók, A. P.; Ortner, C.; Csányi, G.; Ceriotti, M. Physics-Inspired Structural Representations for Molecules and Materials. *Chem. Rev.* **2021**, *121* (16), 9759-9815.

(30) Behler, J. Four Generations of High-Dimensional Neural Network Potentials. *Chem. Rev.* **2021**, *121* (16), 10037-10072.

(31) Meuwly, M. Machine Learning for Chemical Reactions. *Chem. Rev.* **2021**, *121* (16), 10218-10239.

(32) Huang, B.; von Lilienfeld, O. A. Ab Initio Machine Learning in Chemical Compound Space. *Chem. Rev.* **2021**, *121* (16), 10001-10036.

(33) Manzhos, S.; Carrington, T. Neural Network Potential Energy Surfaces for Small Molecules and Reactions. *Chem. Rev.* **2021**, *121* (16), 10187-10217.

(34) Westermayr, J.; Marquetand, P. Machine Learning for Electronically Excited States of Molecules. *Chem. Rev.* **2021**, *121* (16), 9873-9926.

(35) Luo, Y. Chemistry in the Era of Artificial Intelligence. *Prec. Chem.* **2023**, *1* (2), 127-128.

(36) Zhang, Y.; Lin, Q.; Jiang, B. Atomistic neural network representations for chemical dynamics simulations of molecular, condensed phase, and interfacial systems: Efficiency, representability, and generalization. *WIREs Comput. Mol.* **2023**, *13* (3), e1645.

(37) Fu, B.; Zhang, D. H. Accurate fundamental invariant-neural network representation of ab initio potential energy surfaces. *Natl. Sci. Rev.* **2023**, *10* (12).

(38) Xie, X.-T.; Yang, Z.-X.; Chen, D.; Shi, Y.-F.; Kang, P.-L.; Ma, S.; Li, Y.-F.; Shang, C.; Liu, Z.-P. LASP to the Future of Atomic Simulation: Intelligence and Automation. *Prec. Chem.* **2024**.

(39) Zhang, Y.; Xia, J.; Jiang, B. REANN: A PyTorch-based end-to-end multi-functional deep neural network package for molecular, reactive, and periodic systems. *J. Chem. Phys.* **2022**, *156* (11), 114801.

(40) Lee, C.; Yang, W.; Parr, R. G. Development of the Colle-Salvetti correlation-energy formula into a functional of the electron density. *Phys. Rev. B* **1988**, *37*, 785-789.

(41) Frisch, M. J.; Trucks, G. W.; Schlegel, H. B.; Scuseria, G. E.; Robb, M. A.; Cheeseman, J. R.; Scalmani, G.; Barone, V.; Petersson, G. A.; Nakatsuji, H.; et al. Gaussian 16 Rev. B.01. *Wallingford, CT* **2016**.

(42) Dewar, M. J. S.; Thiel, W. Ground states of molecules. 38. The MNDO method. Approximations and parameters. *J. Am. Chem. Soc.* **1977**, *99* (15), 4899-4907.